\documentstyle[aps,eqsecnum]{revtex}
\begin{document}
\twocolumn[\hsize\textwidth\columnwidth\hsize\csname@twocolumnfalse%
\endcsname
\draft

\title{
       Phonon-assisted tunneling
       in asymmetric resonant tunneling structures
}
\author{Jun-jie Shi$^{*,1,3,a}$, 
        Barry C.\ Sanders$^{1,b}$ and Shao-hua Pan$^{1,2,3,c}$}
\address{
        $^1$ Department of Physics, Macquarie University,
        Sydney, New South Wales 2109, Australia \\
        $^2$ Institute of Physics, Chinese Academy of Sciences,
        P.\ O.\ Box 603, Beijing 100080, P.\ R.\ China  \\
        $^3$ China Center of Advanced Science and Technology
        (World Laboratory),
        P.\ O.\ Box 8730, Beijing 100080, P.\ R.\ China\\
	$^*$ On leave from Department of Physics, Henan Normal 
        University, Xinxiang 453002, Henan, P. R. China\\
}
\date{\today}

\maketitle

\begin{abstract}

Based on the dielectric continuum model, we calculated the phonon assisted 
tunneling (PAT) current of general double barrier resoant tunneling 
structures (DBRTSs) including both symmetric and asymmetric ones. The results
indicate that the four higher frequency interface phonon modes (especially the 
one which peaks at either interface of the emitter barrier) dominate the 
PAT processes, which increase the valley
current and decrease the PVR of DBRTSs. We show that an asymmetric structure 
can lead to improved performance.

\end{abstract}

]

\narrowtext

\section{Introduction}
\label{sec:introduction}

The double barrier resonant tunneling structure (DBRTS)\cite{Tsu73}
continues to attract attention both for its potential
applications to electronic devices and 
for its value in exploring fundamental phenomena\cite{Sun98},
including tests of electron-phonon coupling
theories\cite{Turley91,Turley92,Turley94,Mori92,Mains88,Chevoir89,Wingreen89,Roblin93,Cai89,Shi98,Turley93}.
Despite extensive study of the DBRTS,
challenging problems continue to exist
with respect to practical applications and also for modeling the
characteristics of the DBRTS\cite{Sun98}.
Our objective here is to fully
analyze the phonon-assisted tunneling (PAT) current,
especially in the valley region,
and to demonstrate that there can be important advantages in
designing a DBRTS with an asymmetric structure.
We include all of the phonon modes and provide a complete analysis of the
relative importance of these modes,
particularly for the asymmetric DBRTSs where a limited
investigation has been performed to date\cite{Shi98,Turley93}.
It is important to consider all of the phonon modes
when the asymmetric DBRTS is considered\cite{Shi95,Shi96,Shi97}.

The motivations for considering the asymmetric DBRTS are
the possibility for improving performance,
tunability of the negative differential resistance (NDR)\cite{Chen91},
and tuning of charge accumulation in the
quantum well (QW)\cite{Schmidt96,Orellana96}.
This asymmetry can be introduced by varying the material compositions in
the two barriers or by creating two barriers of unequal width,
and we will consider both possibilities here.
Our treatment includes a detailed study of all of the  phonon modes,
assuming the dielectric continuum theory and incorporating the
effects of subband nonparabolicity.

Our main concern is with DBRTS performance.
A large NDR is particularly important for
high-frequency resonant tunneling and rapid switching devices
and is usually quantified by the
peak-to-valley ratio (PVR) of the current-to-voltage characteristic curve.
A large PVR and low valley current are desirable
for most resonant tunneling diodes (RTD) device applications.
Generally, PAT,
$\Gamma -X$ intervalley tunneling, impurity scattering,
the interface roughness scattering and the tunneling of
quasi-two-dimensional subband electron in the pseudo-triangular
well in the emitter can cause the valley current\cite{Roblin93}.
For a polar semiconductor DBRTS, the effects of
phonon scattering on resonant tunneling are very important and
inevitable especially at room temperature. The electrons in the DBRTS
may emit phonons during the resonant tunneling process.
In general, the contribution by the PAT current is
small compared with the coherent resonant tunneling
process but can be large compared to the other contributions to current.
Hence the total tunneling current density can be 
approximated by $J=J_c+J_p$,
for $J_c$ the coherent tunneling current density
and $J_p$ the PAT current density.
Although the PAT current has been investigated 
theoretically\cite{Turley91,Turley92,Turley94,Mori92,Mains88,Chevoir89,Wingreen89,Roblin93,Cai89,Turley93}, 
we provide a more rigorous investigation of electron-phonon scattering
and the PAT current in general DBRTS including both symmetric and asymmetric 
ones, with the contribution of all of the phonon modes
accounted for quantitatively. Hence the PAT physical picture is further 
clarified.

\section{Theory}
\label{sec:theory}

Working within the framework of the dielectric continuum model
and treating the electron-phonon interaction via the Fr\"ohlich-like
Hamiltonian, we can calculate the electron-phonon scattering rate~$W$
according to the Fermi golden rule for two cases\cite{Shi96}:
scattering by interface phonons
$W_{\rm int}^{(i\rightarrow f)}(\vec{k}_i, E_z)$
and by confined LO phonons
$W_{\rm LO}^{(i\rightarrow f)}(\vec{k}_i, E_z)$.
The scattering rate depends explicitly on the phonon occupation
number~$N_{\rm ph}$ which is temperature-dependent according to the Planck
distribution\cite{Shi98}.
As the width of the final resonant state is very narrow,
the final state is treated as a completely localized state in the
well\cite{Turley91,Turley92,Turley94,Chevoir89,Turley93,Vassell83}.
We have recently presented the expressions of  these scattering rates
$W_{\rm int,LO}^{(i\rightarrow f)}(\vec{k}_i, E_z)$
for both symmetric and asymmetric DBRTSs~\cite{Shi98}.

Electron tunneling in a DBRTS depends sensitively on the
bias voltage $V$.
The coherent tunneling current density $J_c$ and PAT current density $J_p$ are given by
\begin{equation}
\label{J_c}
J_c = e n\langle T(E_z)v_z(E_z)\rangle
\end{equation}
and
\begin{equation}
\label{J_p}
J_p=e N\langle W\rangle / A.
\end{equation}
Here $e$ is the absolute value of the electron charge,
$N$ the total electron number in the electron reservoir with volume
$\Omega$, $n=N/\Omega$ the electron density (assumed to be constant),
$T(E_z)$ the electron transmission coefficient,
and~$v_z(E_z)$ the longitudinal electron velocity,
$A$ the cross-sectional area of the structure, and $W$ the
electron-phonon scattering rate\cite{Shi98}. $\langle\rangle$ 
represents averaging on quantum states.

From Eq. ({\ref {J_c}}) we obtain\cite{Tsu73}
\begin{equation}
J_c=J_{c\rightarrow}-J_{c\leftarrow},
\end{equation}
with
\begin{eqnarray}
\label{J_c:rightarrow}
J_{c\rightarrow}
        &=&\frac{e m_{\|}k_BT}{2\pi^2\hbar^3}
        \int_0^\infty T(E_z)    \nonumber       \\
        && \ln\{1+\exp[(E_F-E_z)/k_BT]\} dE_z,
\end{eqnarray}
and
\begin{eqnarray}
\label{J_c:leftarrow}
J_{c\leftarrow}
        &=&\frac{e m_{\|}k_BT}{2\pi^2\hbar^3}
        \int_{0}^\infty T(E_z)  \nonumber       \\
        && \ln\{1+\exp[(E_F-e V-E_z)/k_BT]\} dE_z .
\end{eqnarray}
Here  
$J_{c\rightarrow}$ ($J_{c\leftarrow}$) denotes the tunneling
current density from the emitter (collector) to the collector (emitter), and $J_{c\rightarrow}$ is critical
for the NDR of the DBRTS, which controls the coherent
tunneling current density.
$E_F$ and $E_F-e V$ are, respectively,
the local Fermi energy levels in the
emitter and collector,
and~$m_{\|}$ is the electron effective mass in the $x-y$ plane parallel to
the interfaces of the DBRTS.

From Eq. ({\ref {J_p}}) we derive
\begin{eqnarray}
\label{J_p:solution}
J_p&=&{e m_{\|}k_BTL_e\over
  2\pi^2\hbar^3}\left({m_z\over 2}\right)^{1/2}
        \sum\limits_{p=\nu,m}\int_{E_w+\hbar\omega_p}^\infty E_z^{-1/2}
        \nonumber       \\      &&\times
        \ln\left[ 1+e^{(E_F-E_z)/k_BT} \right]
        W_p(E_z) dE_z,
\end{eqnarray}
with $L_e$ the emitter length, $m_z$ the electron
effective mass for longitudinal motion in the emitter, $\nu=2,3,4$ the index for confined bulk-like LO phonon modes in the
left-barrier, the well
and the right-barrier layers, respectively, 
and $m$ (1 to 8) the index for the eight interface phonon modes.
The electron-phonon scattering
rate in the $\nu^{\rm th}$ layer is
$W_\nu (E_z)$, and $W_m(E_z)$ is the
scattering rate between electron and the 
$m^{\rm th}$ interface phonon mode.
Explicit expresssions for  $W_\nu (E_z)$ and $W_m(E_z)$ 
have been presented in our recent paper\cite{Shi98}.

In obtaining Eq.~(\ref{J_p:solution}) we have assumed ~$W(\vec{k}_i,E_z) \doteq W(0,E_z) \equiv W(E_z)$ 
and have ignored the contribution from
collector-to-emitter tunneling via a phonon absorption\cite{Turley91,Turley92,Turley94}.

Despite the apparent minor difference between Eqs.~(\ref{J_c:rightarrow})
and~(\ref{J_p:solution}),
simulating $J_p$ is
more costly in computer time compared to simulating $J_c$
because $W_p(E_z)$ is more complicated than $T(E_z)$\cite{Shi98}. Our calculations show 
that $J_p$ mainly comes from the electron scattering by the higher frequency 
interface phonons
(especially the interface phonons localized at either interface of the left
barrier).

\section{Numerical results and discussion}

Numerical calculations have been performed for asymmetric and symmetric DBRTSs
${\cal A}(x,d)$ defined as
\begin{eqnarray}
{\cal A}(x,d)
        &\equiv& n^+{\rm GaAs}(1000 \AA)/{\rm Al}_x{\rm Ga}_{1-x}
        {\rm As}(30 \AA)/{\rm GaAs}(60 \AA)  \nonumber       \\
        &&/{\rm Al}_{0.3}
        {\rm Ga}_{0.7}{\rm As}(d)/n^+{\rm GaAs}(1000 \AA) ,
\end{eqnarray}
with $m_{\|}=m_z$ and equal doping concentration  $10^{18}$ cm$^{-3}$ in
emitter and collector. The physical parameters used are the same as in\cite{Shi97}. 

Figure~\ref{fig:dispersion} presents
the dispersion calculated from Eq.~(4) of\cite{Shi96} for the eight interface 
modes
of structure ${\cal A}(0.25,20\AA)$, which reveals that the four lower-frequency modes occupy a much narrower frequency
band than the four higher-frequency modes.
Moreover, the dispersion of the interface modes is significant for the case
$k\leq 0.1 \AA^{-1}$ and negligible for $k>0.1 \AA^{-1}$.
The electron--interface-phonon coupling function $\Gamma (k,z)$\cite{Shi96} is
a complicated function of both $z$ and $k$ in a
DBRTS.
The $\Gamma (k,z)-z$ relation in Fig.~\ref{fig:Gamma-z} reveals the
strength of the electron
interaction with different interface modes peaks at different interfaces.
For example, the electron interaction with mode 7
(denoted e-p(7)) peaks at the $z=0$ and 30 $\AA$
interfaces (i.e., either interface of the left barrier).
We also find that $|\Gamma (k,z)|$
decreases rapidly as a function of $k$ for $0 < k\leq 0.01 \AA^{-1}$, slowly
for $0.01 \AA^{-1}<k<0.08 \AA^{-1}$, and then rapidly 
for $k>0.08 \AA^{-1}$ for the e-p(7) interaction.
The 7$^{\rm th}$ interface mode is much more important than the other modes in
the asymmetric DBRTS,
and the four higher-frequency modes produce intensive
polarisation in the DBRTS, resulting in a significant interaction with 
electrons.
On the contrary, the four lower-frequency modes produce a weak interaction with
electrons compared with the higher-frequency modes, which can be ignored\cite{Shi98,Shi95,Shi96,Shi97}.
In the following calculations,
we will thus only consider the contribution of the four
higher-frequency modes to the scattering rate and the PAT current for
simplicity.

Figure~\ref{fig:W-Ez} shows the scattering rate~$W$
{\em vs} incident electron energy $E_z$ for ${\cal A}(0.25, 20\AA)$.
For $n=10^{18}$ cm$^{-3}$,
the Fermi energy level $E_F=42.5$ meV 
at $T=300$ K.
We observe that the contribution due to interface phonons is larger
than that due to LO bulk-like phonons for the electron with lower incident 
energies.
We know from the Fermi distribution function that
the emitter states are appreciably populated only for
$E_z\leq E_F+k_BT=68.3$ meV (as $T=300$ K).
Hence, we can infer that the interface phonons contribute much more than the
confined
bulk-like LO phonons to the PAT current.

We present the electron-phonon scattering rate, inclouding
the subband nonparabolicity, as the dash-dot-dot line
in Fig.~\ref{fig:W-Ez}, which shows that the subband
nonparabolicity has a large influence on the electron-phonon scattering.
The peak position shifts to a lower energy, and its
value decreases under the influence of the subband nonparabolicity.

In the case of including subband nonparabolicity,
PAT current-to-voltage curves are shown in Fig.~\ref{fig:PAT}
at $T=300$ K for the structure ${\cal A}(0.25,20\AA)$.
Figure~\ref{fig:PAT}(a) shows
the tunneling current assisted by the four higher-frequency interface phonon
modes  and their sum. We can see from
Fig.~\ref{fig:PAT}(a) that the 7$^{\rm th}$ interface mode,
which is localized at either interface of the left barrier,
is the most important of all of the interface modes,
and this result is consistent with the results shown in Fig.~\ref{fig:Gamma-z}.
The total interface PAT current is a complicated function of the applied 
voltage and has two peaks for increasing voltage.
Figure~\ref{fig:PAT}(b) gives the confined bulk-like LO
PAT current density in structure ${\cal A}(0.25,20\AA)$.
This figure indicates that the
PAT current from the LO phonons in the well is
much larger than those from the LO phonons in the two barrier layers
with a complicated behavior for increasing bias voltage.
Figure~\ref{fig:PAT}(c) presents
the total PAT current density including the
interface and the confined bulk-like LO phonons.
We can see from
Fig.~\ref{fig:PAT}(c) that the interface PAT current is
larger by one order of magnitude than the confined LO PAT current, confirming
that interface-phonon scattering dominates over
confined LO phonon scattering (c.f. Fig.~3). Moreover, Fig.~4(c) also shows 
that 
the total PAT current is a very complicated function of the applied voltage 
and has two peaks, similar to the results based on the Green function method\cite{Mori92}. 
This is completely due to the complexity of the contribution of the phonon modes to the PAT current in our DBRTS, as shown in Figure~4(a). 
Figure~\ref{fig:PAT} also indicates that the PAT current is mainly
determined by scattering between electrons and  higher frequency interface
phonons (especially the interface phonons localized at either interface of the
left barrier), showing a clear physical picture for the PAT process in general 
DBRTS.

Figure~\ref{fig:PAT2} shows the total current-to-voltage curve at room
temperature for ${\cal A}(0.25, 20\AA)$,
including coherent and PAT currents.
This figure shows that PAT increases the valley current and
decreases the PVR. The result shown in Fig.~5 is similar to those 
obtained by the Wigner function method\cite{Mains88} and the  Wannier function envelope equation 
method\cite{Roblin93}.

We have also studied the current-voltage characteristics 
for DBRTSs ${\cal A}(0.3, d)$ with $d=20$, $30$ and $40\AA$.
The calculated results show that when the right-barrier thickness $d$ increases,
the peak current decreases, the PVR increases, and the peak position shifts towards 
higher bias voltage. These three characteristics are in agreement with  
recent 
experimental results~\cite{Schmidt96}. Moreover, we have also studied the 
current-voltage characteristics for
${\cal A}(x, 30\AA)$ with $x=0.2$, $0.3$ and $0.4$. The calculated results 
show that when the Al composition $x$, and thus the height of the left barrier,
increases, the peak current decreases, the PVR increases, and the peak position remains at the same bias voltage. These theoretical results, which show that an asymmetric DBRTS can lead to improved performance, await experimental confirmation.

\section{Summary}

Employing the dielectric continuum model,
including all the phonon modes and treating conduction band nonparabolicity, the PAT process is investigated in detail. 
We show that for a DBRTS, no matter it is symmetric or asymmetric one, the four higher frequency modes, which reduce to two symmetric and two antisymmetric higher frequency modes for a symmetric structure, dominate the interface PAT process. In particular, for the interface PAT process in a symmetric DBRTS, the two symmetric higher frequency modes are most important, the two antisymmetric higher frequency modes are less important, and the two symmetric and antisymmetric lower frequency modes are negligible. The above opinion is different from that of Refs.\cite{Turley91,Turley92,Turley94}.  

In general,
the confined LO phonons in the well layer are more important than those
in the two barrier layers, and
the four higher frequency interface modes (i.e., mode 5 to 8, especially the 7$^{\rm th}$
one which peaks at either interface of the left barrier) dominates over the 
four lower frequency interface modes and 
all of the confined LO phonon modes to 
electron-phonon scattering and PAT current.
The PAT increases the valley
current and decreases the PVR of DBRTS. 
It is worth mentioning that the mode 7 becomes the highest frequency symmetric interface mode for a symmetric DBRTS.
The PAT physical picture stated in the above is useful and important for further understanding PAT process in DBRTSs
and for designing better RTD devices.

Subband nonparabolicity has a significant
influence on electron-phonon scattering, PAT, and the
current-to-voltage characteristic of a DBRTS.

We find that the peak current is reduced, the position of peak current is
shifted to
a higher voltage, and the PVR is enlarged if the right-barrier width is
increased when the two barriers have the same height.
The peak current is reduced and the PVR is increased by suitably
increasing the left-barrier height when the two barriers have the same width.
An asymmetric DBRTS with a suitably designed structure has a
larger PVR than its commonly-used symmetric counterpart.
The results obtained in this paper are useful for analysing
coherence-breaking phonon scattering and for potentially important
resonant tunneling device applications.

\acknowledgments

Jun-jie Shi has been supported by an Overseas Postgraduate Research
Scholarship and a Macquarie University International Postgraduate
Research Award.
This work has been supported by an Australian Research
Council Large Grant and by a Macquarie University Research Grant. We
have benefitted from many useful discussions with L.\ Tribe, E.\ M.\ Goldys,
and D.\ J.\ Skellern.

\begin{figure}
\caption{
        The dispersion curves of the interface modes for
        structure ${\cal A}(0.25, 20\AA)$.
        }
\label{fig:dispersion}
\end{figure}
\begin{figure}
\caption{
	Spatial dependence of the 
        normalized coupling functions
        $(\hbar e^2/A\varepsilon_0)^{-1/2} \Gamma(k,z)$
        for the structure ${\cal A}(0.25, 20\AA)$
        ($k=0.01 \AA^{-1}$). Here the numbers by the curves 
	represent the interface-phonon frequency in order of increasing
	magnitude:
        (a) for the four lower-frequency modes and
        (b) for the four higher-frequency modes
        with the interfaces localized at $z=$0, 30, 90 and 110 $\AA$, respectively.
        }
\label{fig:Gamma-z}
\end{figure}
\begin{figure}
\caption{
        The normalized electron-phonon scattering rate~$W/(N_{\rm ph}+1)$
        {\em vs} incident electron energy $E_z$ for ${\cal A}(0.25, 20\AA)$
        at the bias 150 mV.
        The dashed line and dash-dot line represent, respectively,
        the contribution of the interface phonons and confined bulk-like
        LO phonons, and
        the solid line is their sum in the absence of
        subband nonparabolicity.
        The dash-dot-dot line is the total scattering rate including
        subband nonparabolicity.
        }
\label{fig:W-Ez}
\end{figure}
\begin{figure}
\caption{
        PAT current-to-voltage curves at
        room temperature for ${\cal A}(0.25, 20\AA)$, including the
subband
        nonparabolicity: (a) Interface PAT: the
        dashed line represents the 5$^{\rm th}$
        interface mode contribution for the tunneling current,
        the dash-dot-dot line
        for the 6$^{\rm th}$ mode,
        the thin solid line for the 7$^{\rm th}$ mode, the dash-dot
        line for the 8$^{\rm th}$ mode and the heavy solid line is their sum.
        (b) LO PAT: the dashed line represents the
        confined LO phonon in the left barrier
        (Al$_{0.25}$Ga$_{0.75}$As) contribution for the tunneling current, the
        dash-dot line for the LO phonon in the right barrier
(Al$_{0.3}$Ga$_{0.7}$As),
         the thin solid line for the confined bulk-like LO phonon in the well
(GaAs) and the heavy solid line is their
        sum. (c) Interface PAT current (thin
        solid line), confined LO PAT current (dashed
        line) and their sum (heavy solid line).
        }
\label{fig:PAT}
\end{figure}
\begin{figure}
\caption{
        The total current-to-voltage characteristic curve
        at room temperature considering the subband
        nonparabolicity for ${\cal A}(0.25, 20\AA)$.
        The dashed line stands for the coherent
        tunneling current density. The
        solid line is the total tunneling current density combining coherent
        tunneling and
        PAT currents.
        }
\label{fig:PAT2}
\end{figure}
\end{document}